\documentclass[preprint,12pt]{elsarticle}

\usepackage{amssymb}
\usepackage{amsmath}
\usepackage{bm}
\usepackage{mathrsfs}
\usepackage{pifont}    
\usepackage{hyperref} 
\usepackage{booktabs,makecell, multirow, tabularx}
\usepackage{subcaption}
\journal{Information Fusion}

\begin{document}

\begin{frontmatter}



\title{Mixture of Disentangled Experts with Missing Modalities for Robust Multimodal Sentiment Analysis}



\author[1]{Xiang Li}

\author[1]{Xiaoming Zhang\corref{cor1}}
\author[1]{Dezhuang Miao}
\author[2]{Xianfu Cheng}
\author[1]{Dawei Li}
\author[3]{Honggui Han}
\author[2]{Zhoujun Li}
\affiliation[1]{organization={School of Cyber Science and Technology, Beihang University},
    city={Beijing},
    postcode={100191},
    country={China}}
\affiliation[2]{organization={School of Computer Science and Engineering, Beihang University},
    city={Beijing},
    postcode={100191},
    country={China}}
\affiliation[3]{organization={School of Information Science and Technology, Beijing University of Technology},
    city={Beijing},
    postcode={100124},
    country={China}}

\cortext[cor1]{Corresponding author: yolixs@buaa.edu.cn}

\begin{abstract}
Multimodal Sentiment Analysis (MSA) integrates multiple modalities to infer human sentiment, but real-world noise often leads to missing or corrupted data. 
However, existing feature-disentangled methods struggle to handle the internal variations of heterogeneous information under uncertain missingness, making it difficult to learn effective multimodal representations from degraded modalities.
To address this issue, we propose \textbf{DERL}, a  \textbf{D}isentangled \textbf{E}xpert \textbf{R}epresentation \textbf{L}earning framework for robust MSA.  
Specifically, DERL employs hybrid experts to adaptively disentangle multimodal inputs into orthogonal private and shared representation spaces.
A multi-level reconstruction strategy is further developed to provide collaborative supervision, enhancing both the expressiveness and robustness of the learned representations.  
Finally, the disentangled features act as modality experts with distinct roles to generate importance-aware fusion results.
Extensive experiments on two MSA benchmarks demonstrate that DERL outperforms state-of-the-art methods under various missing-modality conditions.  
For instance, our method achieves improvements of 2.47\% in Acc-2 and 2.25\% in MAE on MOSI under intra-modal missingness.
\end{abstract}




\end{frontmatter}



\section{Introduction}
\label{sec1}
Multimodal Sentiment Analysis (MSA) aims to interpret human sentiments by integrating text, visual, and audio modalities \cite{zhu2023multimodal,wang2024cross}. 
Despite the rapid progress in multimodal learning, most existing MSA methods assume that all modalities are fully available during training and inference \cite{yang2022disentangled,zhang2023learning,li2025learning,zhao2025decoupled,wang2025tpin}. 
In real-world scenarios, multimodal data are often incomplete due to sensor failures or background noise, leading to substantial performance degradation for models trained under complete conditions \cite{li2024toward,li2024correlation}.
Recent MSA studies \cite{yuan2021transformer,sun2023efficient,zhang2024towards,guo2024multimodal,zhu2025proxy} have attempted to handle missing modalities.
These methods mainly focus on learning joint multimodal representations~\cite{zhang2024towards,guo2024multimodal} or recovering missing semantics through feature reconstruction~\cite{yuan2021transformer,sun2023efficient}.
For example, LNLN \cite{zhang2024towards} enhances model robustness under incomplete data by taking text as the dominant modality and leveraging others as complementary sources.
However, these methods fail to fully exploit the unique information within each modality. 
Since heterogeneous modalities express sentiment in various ways, they contain both private and shared emotional cues ~\cite{yu2021learning,zeng2024disentanglement}.
Recent efforts have explored disentangled representation learning \cite{yang2022disentangled,li2024toward} to capture heterogeneous sentiment cues and enhance multimodal fusion. For instance, HRLF \cite{li2024toward} factorizes sentiment representations via cross-modal transformation and semantic reconstruction.

Despite the progress, current feature-disentangled methods fail to adapt to heterogeneous information variations arising from internal feature degradation, which occurs under uncertain missingness and manifests in three main challenges.
\hypertarget{one}{\ding{182}} \textit{Inflexible Modality Disentanglement}: Random missingness degrades modality quality and intensity, disrupting the internal balance between private and shared information. Previous static methods~\cite{hazarika2020misa,li2023decoupled} lack the flexibility for adaptive  disentanglement, making it challenging to factorize heterogeneous representations under missingness.
\hypertarget{two}{\ding{183}} \textit{Inadequate Feature Reconstruction}: Auto-regressive reconstruction \cite{hazarika2020misa,wang2025dlf} based solely on degraded disentangled features is under-constraint.  This single-level strategy yields amplified reconstruction errors, suboptimal representations, and insufficient compensation for missing modalities.
\hypertarget{three}{\ding{184}} \textit{Imbalanced Multimodal Fusion}: Different modalities contribute unequally to the fusion results~\cite{zhang2023learning,fang2025emoe}, and random missingness further exacerbates this imbalance. The absence of key sentiment information in certain modalities weakens their contribution in multimodal fusion. A robust fusion strategy needs to account for the relative importance of different modalities under missing scenarios.

To address these issues, we propose a \textbf{D}isentangled \textbf{E}xpert \textbf{R}epresentation \textbf{L}earning (\textbf{DERL}) framework for robust MSA with missing modalities.
DERL first pre-aligns each modality using unified transformer blocks with learnable bottleneck tokens.
Three specialized components are then introduced to handle incomplete modalities: Hybrid Expert Disentanglement (HED), Multi-Level Collaborative Reconstruction (MLCR), and Modality Routing and Fusion (MRF).
First, addressing \hyperlink{one}{{\ding{182}}}, HED adaptively decouples modality-private and shared sentiment features via the mixture of disentangled experts, enabling flexible capture of heterogeneous sentiment cues under missing-modality scenarios.
Second, to compensate for degradation, MLCR reconstructs each modality from multiple views, enforcing semantic richness and providing comprehensive supervision to tackle \hyperlink{two}{{\ding{183}}}. Through collaborative reconstruction, MLCR enhances the quality, completeness, and robustness of the learned representations.
Third, considering \hyperlink{three}{{\ding{184}}}, the disentangled features act as modality experts, and MRF learns the contribution of each expert to enable importance-aware fusion. It emphasizes the interactions between dominant and other modalities to enhance fusion results.
The main contributions of our work are threefold:
\begin{itemize}
    \item DERL introduces a novel integration of disentangled representation learning and mixture-of-experts architecture for robust MSA to address modality information variations under uncertain missingness.
    \item We propose an HED module to disentangle heterogeneous sentiment cues, an MLCR module to recover multi-level degraded representations, and an MRF module to enable importance-aware multimodal fusion.
    \item Extensive experiments on two MSA benchmarks demonstrate that DERL outperforms both feature-disentangled and other strong baselines, showing its robust performance under various missing-modality scenarios.
\end{itemize}
The remainder of this paper is organized as follows. 
Section~\ref{sec:related_work} reviews related work on Robust MSA, disentangled multimodal learning, and mixture-of-experts architecture. 
Section~\ref{sec:method} presents the proposed DERL framework, detailing its mixture of  disentangled experts representation learning mechanism. 
Section~\ref{sec:experiments} describes the experimental setup and reports the results under missing-modality conditions. 
Finally, Section~\ref{sec:conclusion} concludes the paper and discusses potential directions for future research.

\section{Related Work}
\label{sec:related_work}
\subsection{Robust Multimodal Sentiment Analysis}
Multimodal Sentiment Analysis (MSA) integrates sentiment information from multiple modalities for prediction. 
Existing approaches can be broadly categorized into context-based methods and robust noise-aware methods.
Context-based models~\cite{zhang2023learning,zhu2023skeafn,wang2025dlf,li2025multi,zhu2025multimodal}  assume complete multimodal inputs and focus on learning unified representations through cross-modal interactions.
For instance, DMD~\cite{li2023decoupled} decouples multimodal inputs and performs modality-level distillation to enhance representation quality.
However, such models tend to suffer significant degradation when one or more modalities are missing or corrupted.

In contrast, robust noise-aware MSA focuses on real-world scenarios with incomplete or degraded data. Existing solutions can be broadly divided into reconstruction-based methods and joint representation learning methods.
Reconstruction-based approaches \cite{yuan2021transformer, sun2023efficient,li2024correlation} aim to recover missing semantics using complementary cues from available modalities.
For instance, TFR-Net \cite{yuan2021transformer} employs transformer-based reconstruction blocks to restore missing multimodal features, while CorrKD \cite{li2024correlation} enhances cross-modal representations through knowledge distillation.
Joint representation learning methods~\cite{zhang2024towards,yang2026tctr,li2025tf} capture inter-modal correlations to build cohesive shared embeddings.
For example, LNLN~\cite{zhang2024towards} prioritizes textual cues to enhance model robustness by learning language-dominant joint representations.
Recently, feature-disentangled methods~\cite{wang2023multi,li2024toward} have been developed to decouple incomplete multimodal inputs.
However, these methods often overlook the uncertain effects of random noise on internal modality features. 
In this work,  we propose an adaptive disentanglement framework that decouples heterogeneous information from degraded modalities for robust MSA.
\subsection{Disentangled Multimodal Learning}
The inherent cross-modal heterogeneity poses the central challenge for multimodal learning~\cite{yang2022disentangled,zeng2024disentanglement}.
Prior studies have sought to disentangle modality-specific and shared information to obtain comprehensive representations, which can be broadly categorized into space-based methods and mutual information-based  methods.
Space-based methods~\cite{hazarika2020misa,yu2021learning,wang2025dlf} measure similarity and divergence between feature representations to achieve decoupling.
In contrast, mutual information-based approaches~\cite{yang2022disentangled,sun2025multimodal} promote disentanglement by optimizing mutual information bounds or their temporal variants to extract independent yet complementary features across modalities.
Recent works \cite{wang2023multi,li2024toward} extend disentangled learning to handle missing-modality scenarios for robust multimodal analysis.
However, random missing modalities introduce uncertainty that disrupts the proportion between private and shared information, making it difficult to achieve adaptive feature disentanglement under incomplete conditions.

\subsection{Mixture-of-Experts}
Mixture-of-Experts (MoE) models achieve specialization by partitioning network components into multiple expert instances \cite{jordan1994hierarchical}.
The core idea of MoE lies in leveraging cooperation and competition among experts to enhance model capacity.
However, activating all experts simultaneously incurs high computational costs.
Sparse MoE \cite{shazeer2017outrageously} improves efficiency by activating only a subset of experts and has been  adopted in multi-task learning for task-specific expert selection \cite{chen2023adamv} and model compression \cite{li2023merge}.
In MSA, MoE architectures have been employed to replace dense layers for efficiency \cite{zhang2025multimodal} or to handle multiple modalities \cite{fang2025emoe,xu2024leveraging}.
For instance, EMOE \cite{fang2025emoe} formulates each modality as an expert with routing-based cross-modal fusion, while MoMKE \cite{xu2024leveraging} improves modality representations through the integration of unimodal and joint features.
However, most existing MoE-based MSA methods assume complete inputs and fail to preserve modality-specific cues for multimodal fusion.
Differing from prior work, we integrate the MoE architecture into the disentangled representation learning framework, enabling adaptive allocation of modality-private and shared sentiment information from incomplete or noisy multimodal features.
\begin{figure*}[!ht]
\centering
\includegraphics[width=\textwidth]{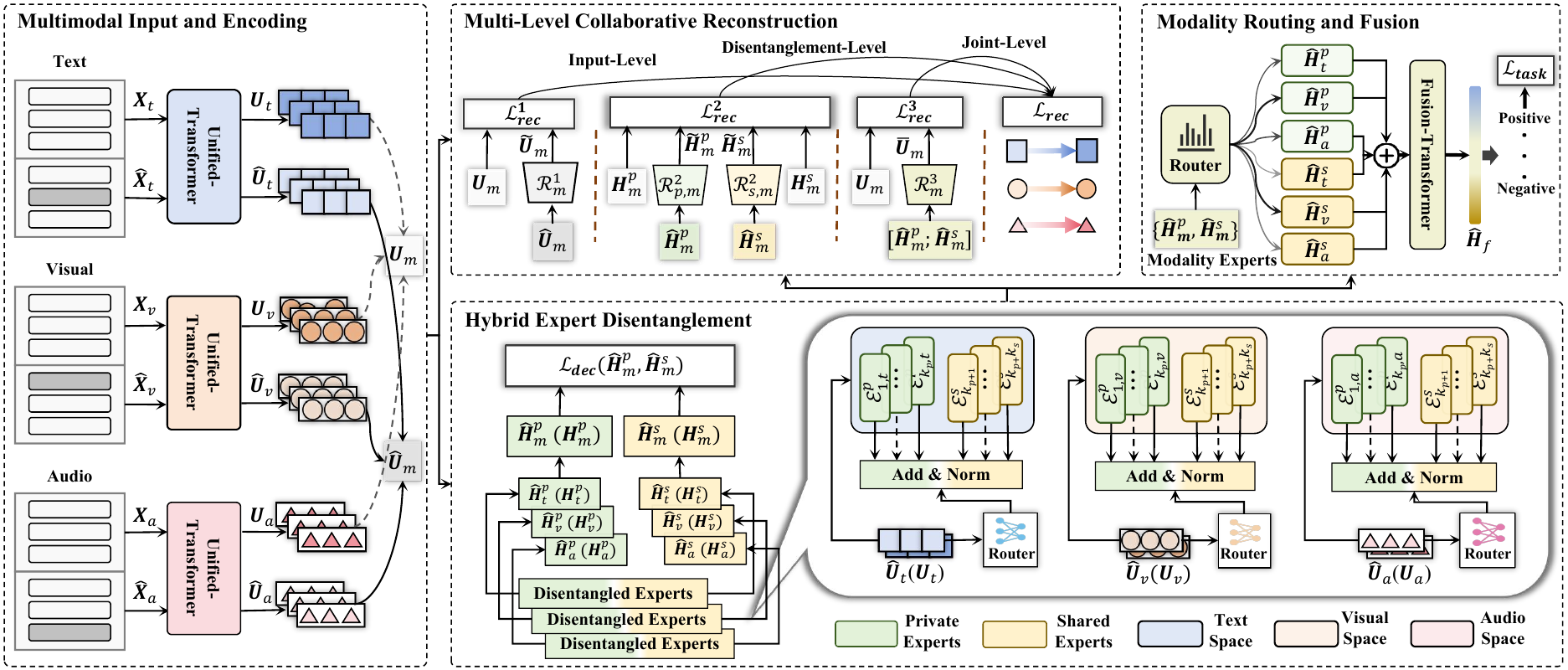}
\caption{
Overview of the proposed DERL framework. White blocks denote complete inputs, while dark gray blocks indicate missing ones.
After multimodal encoding, DERL incorporates three core components to handle incomplete modalities:
\ding{172} Hybrid Expert Disentanglement (HED),
\ding{173} Multi-Level Collaborative Reconstruction (MLCR), and
\ding{174} Modality Routing and Fusion (MRF).
}
\label{fig:Overview}
\end{figure*}
\section{Methodology}
\label{sec:method}
\subsection{Problem Formulation}

We consider three modalities in this work: text ($t$), visual ($v$), and audio ($a$). 
Each modality is represented as a sequence of feature vectors:
$
\bm{X}_t \in \mathbb{R}^{T_t \times D_t}, 
\bm{X}_v \in \mathbb{R}^{T_v \times D_v}, 
\bm{X}_a \in \mathbb{R}^{T_a \times D_a},
$
where $T_m$ denotes the temporal length and $D_m$ represents the feature dimension of modality $m \in \{t, v, a\}$. 
The goal of robust  MSA is to predict the sentiment polarity or intensity $\hat{\bm{Y}}$ of an utterance when uncertain modalities are missing or corrupted. 
Depending on the task setting, the sentiment label $\bm{Y}$ is defined as a continuous scalar representing sentiment intensity ($\bm{Y} \in \mathbb{R}$), which can be discretized into $C$ predefined sentiment categories ($\bm{Y}\in \mathbb{R}^C$) for classification tasks.
\subsection{Framework Overview}
The overall architecture of the proposed DERL framework is illustrated in Figure~\ref{fig:Overview}. 
Given multimodal inputs, DERL first performs the data-missing simulation to generate incomplete samples. 
Unified Transformer-based encoders are then employed to ensure pre-alignment across modalities. 
For each modality, hybrid private and shared disentangled experts are introduced to adaptively factorize modality-specific and shared sentiment representations. 
Subsequently, the reconstruction module collaboratively refines both the original and disentangled features to enhance completeness and robustness.
Finally, an importance-aware fusion module aggregates the disentangled representations through dynamic expert routing for sentiment prediction.
\subsection{Multimodal Input and Encoding}
\paragraph{Random Input Missing}
To emulate real-world multimodal scenarios, we intentionally corrupt the complete modality inputs $\bm{X}_m \in \mathbb{R}^{T_m \times D_m}$.  
Following prior studies~\cite{zhang2024towards}, missing tokens in the text modality are replaced with the special \texttt{[UNK]} token in BERT~\cite{devlin2019bert}, simulating transcription or translation errors.  
For visual and audio modalities, missing frames are substituted with zero vectors to mimic sensor failures or signal degradation.  
Specifically, for each sample, we randomly drop between 0\% and 100\% of features for each modality independently:
\begin{equation}
\{\hat{\bm{X}}_t, \hat{\bm{X}}_v, \hat{\bm{X}}_a\} = \text{Random-Missing}(\{\bm{X}_t, \bm{X}_v, \bm{X}_a\}),
\end{equation}
where $\hat{\bm{X}}_m$ denotes the corrupted sequences of modality $m$.
\paragraph{Unified Transformer Encoding}
To standardize multimodal features, we first apply linear projection $\mathcal{P}$ to unify feature dimensions, followed by Transformer encoders to capture sequential dependencies.  
A set of learnable shared bottleneck tokens $\bm{U}_b \in \mathbb{R}^{N \times D}$ is introduced as the bridge for pre-alignment. Through self-attention, these tokens absorb consistent cues and embed them into common spaces.  
The encoded incomplete features are computed as:
\begin{equation}
\hat{\bm{U}}_m = \text{Unified-Transformer}_m([\bm{U}_b; \mathcal{P}_m(\hat{\bm{X}}_m)]),
\end{equation}
where $[\cdot; \cdot]$ denotes concatenation, $\hat{\bm{U}}_m \in \mathbb{R}^{N \times D}$, and $D$ denotes the common feature dimension of the unified representation for modality $m$.
Similarly, the complete modality inputs $\bm{X}_m$ are  encoded to obtain unified representations $\bm{U}_m = \{\bm{U}_t, \bm{U}_v, \bm{U}_a\}$ for subsequent processing.

\subsection{Hybrid Expert Disentanglement}
Given the uncertainty of modality degradation, the internal heterogeneous information within each modality varies in a non-quantifiable manner.  
Inspired by MoE, the proposed {Hybrid Expert Disentanglement (HED)} module adaptively allocates inputs to learnable {private} and {shared} experts, enabling data-driven disentanglement based on the observed degree of modality missingness.

HED maintains three sets of hybrid disentangled experts:
\[
\{ \mathcal{E}_{1,m}^{p}, \mathcal{E}_{2,m}^{p}, \dots, \mathcal{E}_{k_p,m}^{p},\mathcal{E}_{k_p+1}^{s}, \mathcal{E}_{k_p+2}^{s}, \dots, \mathcal{E}_{k_p+k_s}^{s} \}.
\]  
The private experts $\{\mathcal{E}^p_{k,m}\}_{k=1}^{k_p}$ capture modality-specific cues and distributional characteristics, 
while the shared experts $\{\mathcal{E}^s_k\}_{k=k_p+1}^{k_p+k_s}$ learn modality-invariant parts consistent across modalities.
Each expert is implemented as a two-layer MLP, with shared experts sharing parameters across all modalities. The soft routing networks $\mathcal{G}_m^e$ then produce token-wise gating scores.
For each token, the routers output the $(k_p + k_s)$-dimensional score vectors, which are normalized using a temperature-scaled Softmax:
\begin{equation}
\bm{W}^{e}_m = \sigma ( \frac{\mathcal{G}_m^e(\hat{\bm{U}}_m)}{\tau_e}),
\end{equation}
where $\sigma(\cdot)$ denotes Softmax function, and $\tau_e$ is a learnable temperature, initialized as $\tau = 1/(k_p + k_s)$ to control the sharpness of expert assignments.
The private and shared representations are obtained via hybrid expert outputs:
\begin{equation}
\begin{aligned}
\hat{\bm{H}}_m^p &= \sum_{k=1}^{k_p} \bm{w}_{m,k}^e\, \mathcal{E}^p_{k,m}(\hat{\bm{U}}_m),\\
\hat{\bm{H}}_m^s &= \sum_{k=k_p+1}^{k_p+k_s} \bm{w}_{m,k}^e\, \mathcal{E}^s_{k}(\hat{\bm{U}}_m),
\end{aligned}
\label{eq:HAD_outputs}
\end{equation}
where $\bm{w}_{m,k}^e$ denotes the routing weight for the $k$-th expert.

For complete inputs $\bm{U}_m$, the same HED encoder is applied to ensure consistent feature decomposition, producing decoupled features $\bm{H}_m^p$ and $\bm{H}_m^s$.  
To enhance the discriminability of disentangled representations, we impose the soft orthogonality constraint that encourages private and shared features to reduce redundancy across subspaces:
\begin{equation}
\mathcal{L}_{dec} = \sum_{m \in \{t, v, a\}} \cos(\hat{\bm{H}}_m^p, \hat{\bm{H}}_m^s),
\end{equation}
where $\cos(\cdot, \cdot)$ denotes the cosine similarity.

\subsection{Multi-Level Collaborative Reconstruction}
To enhance the quality of modality representations under incomplete inputs, we propose the Multi-Level Collaborative Reconstruction (MLCR) mechanism.
MLCR performs feature reconstruction at three levels: input, disentanglement, and joint. It recovers degraded semantics through multi-view supervision, thereby improving the robustness of representations to missing or noisy modalities.
\paragraph{Input-Level Reconstruction}  
The first level aims to reconstruct the complete features from corrupted inputs:
\begin{equation}
\tilde{\bm{U}}_m = \mathcal{R}^1_m(\hat{\bm{U}}_m),
\end{equation}
where $\mathcal{R}^1_m(\cdot)$ denotes the first-level reconstruction network, e.g., a two-layer MLP.  
The corresponding reconstruction loss is defined using the L1 norm:
\begin{equation}
\mathcal{L}_{rec}^1 = \sum_{m \in \{t,v,a\}} \| \tilde{\bm{U}}_m - \bm{U}_m \|_1.
\end{equation}

\paragraph{Disentanglement-Level Reconstruction}  
At the second level, the private and shared disentangled representations of each incomplete modality are reconstructed to align with their complete counterparts, respectively:
\begin{equation}
\tilde{\bm{H}}_m^p = \mathcal{R}_{p,m}^2(\hat{\bm{H}}_m^p), \:   
\tilde{\bm{H}}_m^s = \mathcal{R}_{s,m}^2(\hat{\bm{H}}_m^s),
\end{equation}
where $\mathcal{R}_{p,m}^2$ and $\mathcal{R}_{s,m}^2$ are independent reconstruction networks for private and shared parts. 
The corresponding reconstruction loss is calculated as follows:
\begin{equation}
\mathcal{L}_{rec}^2 = \sum_{m \in \{t,v,a\}} ( \| \tilde{\bm{H}}_m^p - \bm{H}_m^p \|_1 + \| \tilde{\bm{H}}_m^s - \bm{H}_m^s \|_1 ).
\end{equation}

\paragraph{Joint-Level Reconstruction}  
Finally, the disentangled features from missing modalities are combined to jointly reconstruct the encoded complete features:
\begin{equation}
\bar{\bm{U}}_m = \mathcal{R}^3_m([\hat{\bm{H}}_m^p; \hat{\bm{H}}_m^s]).
\end{equation}
Here $\mathcal{R}^3_m(\cdot)$ denotes the joint-level reconstruction network.
The reconstruction loss at this level is:
\begin{equation}
\mathcal{L}_{rec}^3 = \sum_{m \in \{t,v,a\}} \| \bar{\bm{U}}_m - \bm{U}_m \|_1.
\end{equation}
The overall reconstruction loss combines all three levels:
\begin{equation}
\mathcal{L}_{rec} = \frac{1}{3} \left( \mathcal{L}_{rec}^1 + \mathcal{L}_{rec}^2 + \mathcal{L}_{rec}^3 \right).
\end{equation}

\subsection{Modality Routing and Fusion}
Under incomplete conditions, the relative importance of each modality becomes more variable. To this end, a robust fusion strategy should emphasize the dominant modality while leveraging other modalities. Accordingly, we propose the Modality Routing and Fusion (MRF) module, which adaptively identifies the dominant modality and assigns appropriate weights to each modality during fusion.

Given  $\hat{\bm{H}}_m^p$ and $\hat{\bm{H}}_m^s$,  
each modality acts as a distinct expert, allowing the model to emphasize stronger modalities while leveraging complementary cues from others.  
The features are first combined to form an expert sequence:
\begin{equation}
{\hat{\bm{E}}}_c = \{\hat{\bm{H}}^p_t, \hat{\bm{H}}^p_v,\hat{\bm{H}}^p_a, \hat{\bm{H}}^s_t, \hat{\bm{H}}^s_v,\hat{\bm{H}}^s_a\}.
\end{equation}

Next, a two-layer MLP routing network $\mathcal{G}^r$ computes token-wise routing weights for all experts:
\begin{equation}
\bm{W}^r = \sigma ( \frac{\mathcal{G}^r(\hat{{\bm{E}}}_c)}{\tau_r}) 
= [\bm{w}_t^p, \bm{w}_v^p, \bm{w}_a^p, \bm{w}_t^s, \bm{w}_v^s, \bm{w}_a^s],
\end{equation}
where $\tau_r$ is initialized as $1 / k_e$ with $k_e = 6$, corresponding to the private and shared components of the three modalities.
The enhanced multimodal sequence is then obtained through weighted aggregation:
\begin{equation}
\hat{\bm{H}}_e = 
\sum_{m \in \{t, v, a\}} \bm{w}_m^p \hat{\bm{H}}_m^p
+ \sum_{m \in \{t, v, a\}} \bm{w}_m^s \hat{\bm{H}}_m^s.
\end{equation}
This adaptive weighting allows the model to emphasize the dominant modalities in the expert sequence.

Finally, Transformer-based fusion layers are employed to capture deeper cross-modal interactions:
\begin{equation}
\hat{\bm{H}}_f = \text{Fusion-Transformer}(\hat{\bm{H}}_e),
\end{equation}
where $\hat{\bm{H}}_f$ denotes the final fusion results, which is robust to imbalanced modalities under uncertain missingness.

\subsection{Objective Optimization}
For task learning, mean pooling is applied to the fused representation $\hat{\bm{H}}_f$, which is fed into an MLP to produce the regression score $\hat{\bm{Y}}$.
The training objective minimizes the mean squared error (MSE) between the predicted score $\hat{\bm{Y}}$ and the ground truth ${\bm{Y}}$ over a batch of size $B$:
\begin{equation}
\mathcal{L}_{{task}} = \frac{1}{B}\sum_{i=1}^{B} \|\hat{\bm{Y}}_i - \bm{Y}_i\|^2_2,
\end{equation}

The overall objective combines three complementary terms:
the regression task loss $\mathcal{L}_{{task}}$,
the disentanglement regularization loss $\mathcal{L}_{{dec}}$,
and the  reconstruction loss $\mathcal{L}_{{rec}}$:
\begin{equation}
\mathcal{L} = \mathcal{L}_{{task}} + \mathcal{L}_{{dec}} + \mathcal{L}_{{rec}}.
\end{equation}

\section{Experiments}
\label{sec:experiments}
\subsection{Datasets and Evaluation Metrics}
\paragraph{Datasets}
We conduct experiments on two widely used MSA benchmarks: MOSI~\cite{zadeh2016multimodal} and MOSEI~\cite{zadeh2018multimodal}.
\textbf{MOSI} comprises 2,199 YouTube utterances annotated with sentiment scores on a 7-point scale from $-3$ (strongly negative) to $+3$ (strongly positive), split into 1,284 training, 229 validation, and 686 test samples.
\textbf{MOSEI} extends MOSI with 22,856 annotated video clips from over 1,000 speakers across approximately 250 topics. It contains 16,326 training, 1,871 validation, and 4,659 test samples, with sentiment scores labeled on the same 7-point scale.
Further details on feature extraction and data processing are provided in \ref{sec:app1}.
\paragraph{Evaluation Metrics}
Following prior works~\cite{sun2023efficient, zhang2024towards}, we evaluate model robustness under  intra-modal missing conditions on regression and classification tasks.
For regression, we adopt the Mean Absolute Error (MAE) and the Pearson Correlation Coefficient (Corr) to measure prediction deviation and correlation with the ground truth.
For classification, continuous sentiment scores are discretized into 7, 5, and 2 classes.
For binary classification, we report Acc-2 and F1 under two configurations: negative vs.\ non-negative (left of “/”) and negative vs.\ positive (right of “/”).
Except for MAE, higher values indicate better performance.
For inter-modal missing experiments, we follow~\cite{zhu2025proxy} and use the F1 score as the evaluation metric.
\subsection{Implementation Details}
All experiments are implemented using the PyTorch framework. The AdamW optimizer with a cosine annealing learning rate schedule is employed for model training. Features from all modalities are aligned in the 128-dimensional space. The initial learning rate is set to $1e^{-4}$, with a batch size of 64 and a total of 200 training epochs. All hyperparameters are determined via grid search. For both MOSI and MOSEI, the shared bottleneck lengths are set to 4 and 8, respectively. The model employs \{1, 3\} private and \{3, 3\} shared experts, and the unified/fusion transformer blocks are configured with depths of \{1, 1\} and \{2, 2\}.

For fair comparison, we follow the missing-modality protocol in LNLN~\cite{zhang2024towards}. During training, 50\% samples in each epoch are randomly selected, and modality features are randomly dropped at rates between 0\% and 100\% to simulate missing-modality conditions. The model tested with the missing rate $r=0.5$ is selected for evaluation. 
\begin{table}[!h]
\centering
\caption{Performance comparison on MOSI and MOSEI under intra-modal missing settings, averaged across all missing rates. Results marked with $\ast$ are reproduced from public codebases under identical settings, while others are reported from ~\cite{zhang2024towards,li2025tf}.}
\resizebox{\textwidth}{!}{%
\begin{tabular}{ccccccc}
\toprule
Method & MAE & Corr & Acc-2 & F1 & Acc-5 & Acc-7 \\
\midrule
\multicolumn{7}{c}{\textbf{MOSI}} \\
\midrule
MISA      & 1.085 & 0.524 & 70.33 / 71.49 & 70.00 / 71.28 & 33.08 & 29.85 \\
Self-MM   & 1.070 & 0.512 & 69.26 / 70.51 & 67.54 / 66.60 & 34.67 & 29.55 \\
MMIM      & 1.077 & 0.507 & 67.06 / 69.14 & 64.04 / 66.65 & 33.77 & 31.30 \\
TFR-Net   & 1.200 & 0.459 & 66.35 / 68.15 & 60.06 / 61.73 & 34.67 & 29.54 \\
CENET     & 1.080 & 0.504 & 67.73 / 71.46 & 64.85 / 68.41 & 33.62 & 30.38 \\
ALMT      & 1.083 & 0.498 & 68.39 / 70.40 & 71.80 / 72.57 & 33.42 & 30.30 \\
DMD$^\ast$       & 1.097 & 0.511 & 71.11 / 71.89 & 71.15 / 72.03 & 37.14 & 33.25 \\
LNLN      & 1.062 & 0.503 & 70.11 / 71.91 & 70.02 / 71.71 & 36.25 & 32.53 \\
EMOE$^\ast$      & 1.114 & 0.513 & 70.92 / 71.65 & 70.69 / 71.71 & 36.55 & 32.90 \\
\textbf{DERL} 
          & \textbf{1.038} & \textbf{0.530} 
          & \textbf{72.19} / \textbf{73.69} 
          & \textbf{71.95} / \textbf{73.57} 
          & \textbf{39.48} & \textbf{34.50} \\
\midrule
\multicolumn{7}{c}{\textbf{MOSEI}} \\
\midrule
MISA      & 0.780 & 0.503 & 75.82 / 71.27 & 68.73 / 63.85 & 39.39 & 40.84 \\
Self-MM   & 0.695 & 0.498 & 77.42 / 73.89 & 72.31 / 68.92 & 45.38 & 44.70 \\
MMIM      & 0.739 & 0.489 & 75.89 / 73.32 & 70.32 / 68.72 & 41.74 & 40.75 \\
TFR-Net   & 0.697 & 0.489 & 77.23 / 73.62 & 71.99 / 68.80 & 34.67 & 46.83 \\
CENET     & 0.685 & 0.535 & 77.34 / 74.67 & 74.08 / 70.68 & \textbf{47.83} & \textbf{47.18} \\
ALMT      & 0.674 & 0.481 & 77.54 / 76.64 & 78.03 / 77.14 & 41.64 & 40.92 \\
DMD$^\ast$       & 0.679 & 0.566 & 75.33 / 77.48 & 76.04 / 77.08 & 47.19 & 46.70 \\
LNLN      & 0.692 & 0.530 
          & {78.19} / 76.30 
          & \textbf{79.95} / {77.77} 
          & 46.17 & 45.42 \\
EMOE$^\ast$      & 0.679 & 0.565 & 73.88 / 76.65 & 74.42 / 76.36 & 45.57 & 46.16 \\

\textbf{DERL} 
          & \textbf{0.665} & \textbf{0.585} 
          & \textbf{78.60} / \textbf{78.16} 
          & 77.43 / \textbf{77.82} 
          &  47.64 &  {46.84}\\
\bottomrule
\end{tabular}
\label{tab:robustness_derl_mosi_mosei}
}
\end{table}

\subsection{Baseline Models}\label{sec:main}
We compare DERL with several representative state-of-the-art methods, including both the complete-modality and missing-modality approaches. The former methods include MISA~\cite{hazarika2020misa}, Self-MM~\cite{yu2021learning}, MMIM~\cite{han2021improving}, CENET~\cite{wang2022cross}, CubeMLP~\cite{sun2022cubemlp}, ALMT~\cite{zhang2023learning}, DMD~\cite{li2023decoupled}, and EMOE~\cite{fang2025emoe}. The latter baselines include MCTN~\cite{pham2019found}, TransM~\cite{wang2020transmodality}, TFR-Net~\cite{yuan2021transformer}, SMIL~\cite{ma2021smil}, GCNet~\cite{lian2023gcnet}, CorrKD~\cite{li2024correlation}, and LNLN~\cite{zhang2024towards}. Comprehensive experiments are conducted to evaluate the  robustness of DERL under intra-modal and inter-modal missing conditions. Detailed descriptions are provided in \ref{sec:app2}.
\begin{figure}[t]
	\centering
	\includegraphics[width=\linewidth]{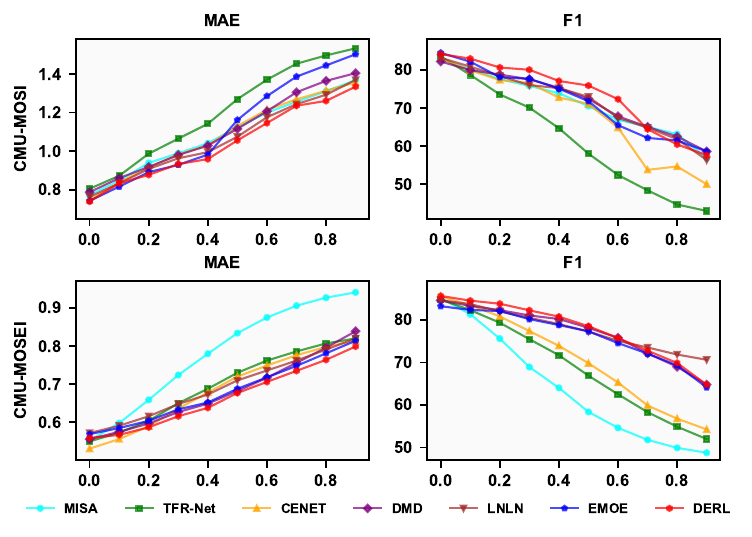}
\caption{Performance trends of representative models across varying missing rates on MOSI and MOSEI datasets.}
	\label{fig:robustness_details}
\end{figure}
\subsection{Robustness Comparison}
\paragraph{Results on Intra-Modal Missingness}  
Following~\cite{zhang2024towards},  we evaluate DERL under intra-modal missing conditions by randomly masking each modality at rates $r \in [0, 0.9]$ with 0.1 increments. The extreme case $r = 1.0$, where all modalities are absent, is excluded as ill-posed. Performance is averaged over all missing rates to assess robustness under varying noise levels.
The results on MOSI and MOSEI are summarized in Table~\ref{tab:robustness_derl_mosi_mosei}, while Figure~\ref{fig:robustness_details} illustrates performance trends under varying missing rates. Overall, DERL outperforms existing methods, demonstrating superior noise robustness and achieving the best competitive results across most metrics.  These results reveal three key observations:
(i) Model performance declines as missing rates increase, indicating that the loss of intra-modal information weakens the expressive power of multimodal features and the quality of joint embeddings.  
(ii) Compared to complete-modality methods such as MISA, DMD, and EMOE, DERL demonstrates stronger robustness under intra-modal missingness. This advantage arises from its adaptive disentanglement via the mixture-of-experts, which effectively captures heterogeneous sentiment cues from incomplete inputs.  
(iii) Against state-of-the-art missing-modality methods like LNLN, DERL achieves notable gains, improving Acc-2 by 2.47\% on MOSI and 2.43\% on MOSEI, while reducing regression errors by 2.25\% and 3.90\%, respectively. These results stem from DERL’s ability to adaptively disentangle and integrate heterogeneous information under varying modality degradation, while multi-level reconstruction enhances the recovery of critical semantics from incomplete inputs. 
\begin{table}[t]
    \centering
    \caption{Performance under inter-modal missing and complete-modality testing conditions on MOSI and MOSEI. Results for baselines are from~\cite{li2024correlation} and $\{m\}$ denotes the available modalities.}
    \resizebox{\linewidth}{!}{
    \begin{tabular}{ccccccccc}
        \toprule   
        {Method} & \{t\} & \{v\} & \{a\} & \{t,v\} & \{t,a\} & \{v,a\} & \{t,v,a\} & Avg. \\
        \midrule
        \multicolumn{9}{c}{MOSI} \\
        \midrule
        Self-MM     & 67.80 & 38.52 & 40.95 & 74.97 & 69.81 & 47.12 & \textbf{84.64} & 56.53 \\
        CubeMLP     & 64.15 & 43.24 & 38.91 & 65.12 & 63.76 & 47.92 & 84.57 & 53.85 \\
        DMD         & 68.97 & 42.26 & 43.33 & 68.45 & 70.51 & 50.47 & 84.50 & 57.33 \\
        MCTN        & 75.21 & 58.57 & 59.25 & 74.82 & 77.81 & 64.21 & 80.12 & 68.31 \\
        TransM      & 77.64 & 56.48 & 63.57 & 80.90 & 82.07 & 67.24 & 82.57 & 71.32 \\
        SMIL        & 78.26 & 59.67 & 67.69 & 79.15 & 79.82 & 71.24 & 82.85 & 72.64 \\
        GCNet       & 80.91 & 58.70 & 65.07 & 83.58 & \textbf{84.73} & 70.02 & 83.20 & 73.84 \\
        CorrKD & 81.20 & \textbf{60.72} & \textbf{66.52} & 82.41 & 83.56 & \textbf{73.74} & 83.94 & \textbf{74.69} \\       
        \textbf{DERL} & \textbf{84.18} & 50.52 & 50.52 & \textbf{84.32} & {84.18} & 50.52 & 84.16 & 69.77 \\
        \midrule
        \multicolumn{9}{c}{MOSEI} \\
        \midrule
        Self-MM     & 71.53 & 37.61 & 43.57 & 74.62 & 75.91 & 49.52 & 83.69 & 58.79 \\
        CubeMLP     & 67.52 & 32.58 & 39.54 & 70.06 & 71.69 & 48.54 & 83.17 & 54.99 \\
        DMD         & 70.26 & 39.84 & 46.18 & 72.45 & 74.78 & 52.70 & 84.78 & 59.37 \\
        MCTN        & 75.50 & 59.46 & 62.72 & 77.13 & 76.64 & 64.84 & 81.75 & 69.38 \\
        TransM      & 77.98 & 58.67 & 63.68 & 78.61 & 80.46 & 62.24 & 81.48 & 70.27 \\
        SMIL        & 76.57 & 60.57 & 65.96 & 76.24 & 77.68 & 66.87 & 80.74 & 70.65 \\
        GCNet       & 80.52 & 61.83 & \textbf{66.54} & 81.15 & 81.96 & 69.21 & 82.35 & 73.54 \\
        {CorrKD} & 80.76 & \textbf{62.30} & 66.09 & 81.28 & 81.74 & \textbf{71.92} & 82.16 & 74.02 \\
        \textbf{DERL} & \textbf{85.19} & 60.46 & 60.41 & \textbf{85.57} & \textbf{85.16} & 60.08 & \textbf{85.57} & \textbf{74.63} \\
        \bottomrule
    \end{tabular}
    }
    \label{tab:Missing_MOSI_MOSEI}
\end{table}
\paragraph{Results on Inter-Modal Missingness}  
Following~\cite{zhu2025proxy}, we evaluate DERL using F1 scores under inter-modal missing conditions by removing one or more modalities. Results on MOSI and MOSEI are shown in Tables~\ref{tab:Missing_MOSI_MOSEI}, with “Avg.” reporting the mean performance across six missing configurations. We have the following key observations:
(i) Inter-modal missing generally reduces model performance, highlighting the importance of leveraging cross-modal cues to enhance fused representations. Models that include the text modality consistently outperform other combinations, confirming that textual information conveys richer and more direct sentiment cues.
(ii) DERL demonstrates superior robustness under inter-modal missingness compared to most baselines. On MOSI, it improves average F1 by 21.6 percentage points over the comparable disentangled method DMD. On MOSEI, DERL achieves an average F1 gain of 0.82\% over the next-best model CorrKD.
(iii) While methods trained under missing-modality conditions perform well in robustness tests and complete-modality trained models excel on complete inputs, DERL achieves both strong robustness and competitive complete-modality performance. For example, on complete-modality MOSEI evaluation, DERL reaches 85.57\% of the best-reported F1, highlighting the effectiveness of the MoE-based disentanglement framework.
(iv) We also observe that under high-noise conditions, DERL converges to similar results across different missing configurations. This phenomenon likely stems from lazy prediction, as increased noise hinders the learning and decoupling of meaningful sentiment information. Moreover, models trained solely under intra-modal missing conditions may fail to fully generalize to inter-modal missing scenarios. Enhancing model robustness across diverse missing conditions remains a key challenge in robust MSA.

\begin{table}[h]
\centering
\caption{Ablation analysis of different components and strategies under intra-modal missingness on MOSI.}
\resizebox{\linewidth}{!}{
\begin{tabular}{ccccccccc}
\toprule
Model  & MAE & Corr & Acc-2 & F1 & Acc-5 & Acc-7 \\
\midrule
w/o HED & 1.065 & 0.518 & 71.14 / 72.13 & 70.97 / 72.07 & 37.39 & 33.62 \\
\midrule
w/o MLCR & 1.073 & 0.506 & 70.13 / 71.65 & 69.94 / 71.58 & 37.89 & 33.88 \\
w/ $\mathcal{L}_{rec}^1$   & 1.061 & 0.510 & 70.99 / 71.84 & 71.02 / 71.78 & 35.95 & 31.80 \\  
w/ $\mathcal{L}_{rec}^2$   & 1.070 & 0.501 & 71.07 / 71.75 & 70.92 / 71.60 & 38.75 & \textbf{34.78} \\  
w/ $\mathcal{L}_{rec}^3$   & 1.055 & 0.523 & 70.41 / 72.91 & 70.29 / 72.11 & 36.72 & 32.30 \\  
\midrule
w/o MRF & 1.066 & 0.497 & 70.66 / 71.45 & 70.69 / 71.58 & 37.22 & 33.64 \\
\midrule
\textbf{DERL} & \textbf{1.038} & \textbf{0.530} & \textbf{72.19} / \textbf{73.69} & \textbf{71.95} / \textbf{73.57} & \textbf{39.48} & {34.50} \\
\bottomrule
\end{tabular}
}
\label{tab:Intra-Ablation}
\end{table}
\begin{table}[h]
    \centering
    \caption{Ablation analysis of different components and strategies under inter-modal missingness on MOSI.}
    \resizebox{\linewidth}{!}{
    \begin{tabular}{ccccccccc}
        \toprule   
        {Model} & \{t\} & \{v\} & \{a\} & \{t,v\} & \{t,a\} & \{a,v\} & \{t,v,a\} & Avg. \\
        \midrule
        w/o HED    & 83.82 & {33.87} & {33.00} & 83.82 & 83.82 & {33.29} & 83.82 & {62.20} \\
        \midrule
        w/o MLCR     & 83.34 & 48.16 & 48.16 & 83.19 & 83.34 & 48.16 & 83.34 & 68.24 \\
        w/ $\mathcal{L}_{rec}^1$         & 84.07 & 37.45 & 37.45 & 84.07 & 84.07 & 37.45 & 84.07 & 64.09 \\
        w/ $\mathcal{L}_{rec}^2$        & 82.15 & 42.26 & 38.14 & 82.47 & 82.28 & 41.61 & 83.04 & 64.56 \\
        w/ $\mathcal{L}_{rec}^3$      & 83.43 & 42.31 & 42.31 & 83.43 & 83.43 & 42.31 & 83.43 & 65.81\\
        \midrule
        w/o MRF        & 83.30 & 48.16 & 48.16 & 83.30 & 83.30 & 48.16 & 83.30 & 68.24 \\
        \midrule
        \textbf{DERL} & \textbf{84.18} & \textbf{50.52} & \textbf{50.52} & \textbf{84.32} & \textbf{84.18} & \textbf{50.52} &  \textbf{84.16}& \textbf{69.77} \\
        \bottomrule
    \end{tabular}
    }
    \label{tab:Inter-Ablation}
\end{table}
 
\subsection{Ablation Study}\label{sec:ablation}

We conduct a thorough ablation study on MOSI to investigate the contribution of each key component in DERL under both intra-modal and inter-modal missing scenarios. The ablation results are summarized in Tables~\ref{tab:Intra-Ablation} and~\ref{tab:Inter-Ablation}, and reveal the following insights:
(i) The removal of the HED module causes consistent performance degradation across missing scenarios, demonstrating the critical role of adaptive expert disentanglement in effectively managing incomplete modalities.
(ii) Similarly, omitting the MLCR module leads to a clear performance decline, showing that feature reconstruction is crucial for maintaining robustness under missing conditions. In addition, using single-level reconstruction alone yields suboptimal results, highlighting the value of multi-level collaborative reconstruction for better feature recovery and semantic richness.
(iii) Excluding the MRF mechanism leads to a noticeable performance drop, highlighting the necessity of importance-aware fusion  under missing-modality conditions.
The full DERL model attains the best performance, demonstrating the effectiveness and robustness of the proposed components.

\begin{table}[t]
    \centering
    \caption{Results with various predefined missing rates on MOSI.}
    \resizebox{\linewidth}{!}{
    \begin{tabular}{ccccccc}
        \toprule   
        {Method} & MAE & Corr & Acc-2 & F1 & Acc-5 & Acc-7 \\
        \midrule
        $r=0.1$  & 1.099 & 0.481 & 70.89/72.35 & 70.73/72.14 & 38.80 & 34.23 \\
        $r=0.3$  & \textbf{1.038} & \textbf{0.530} & 71.63/73.40 & 71.17/73.08 & 38.80 & 34.23 \\
        $r=0.5$  & \textbf{1.038} & \textbf{0.530} & \textbf{72.19}/\textbf{73.69} & \textbf{71.95}/\textbf{73.57} & \textbf{39.48} & {34.50} \\
        $r=0.7$  & 1.054 & 0.511 & 71.52/72.79 & 71.44/72.38 & \textbf{39.48} & \textbf{34.78} \\
        $r=0.9$  & 1.054 & 0.512 & 70.47/73.14 & 70.51/72.98 & 36.49 & 33.78 \\
        \bottomrule
    \end{tabular}
    }
    \label{tab:missing_rate_sensitivity}
\end{table}
\begin{figure}[t]
    \centering
    \begin{subfigure}[b]{\textwidth}
        \centering
        \includegraphics[width=\textwidth]{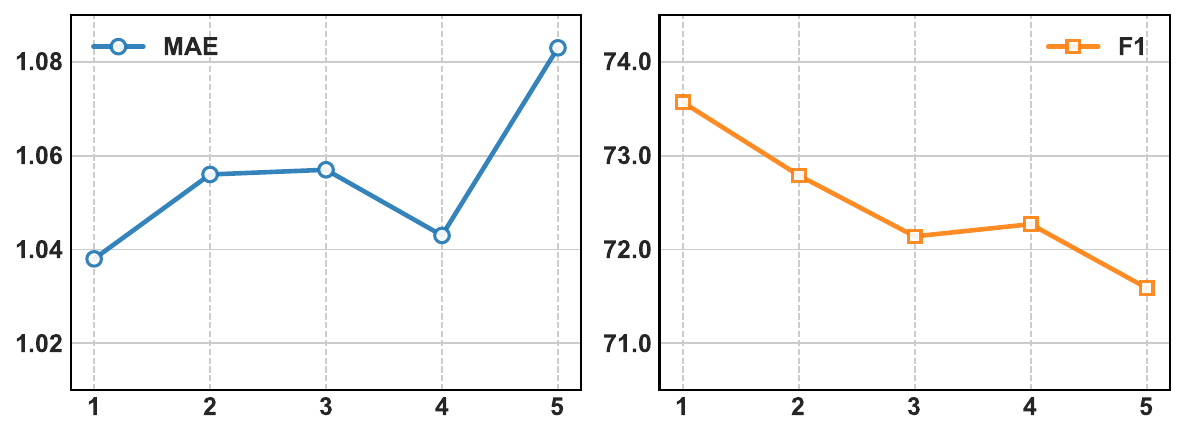}
        \caption{Effect of private experts}
        \label{fig:sub1}
    \end{subfigure}
    \hfill
    \begin{subfigure}[b]{\textwidth}
        \centering
        \includegraphics[width=\textwidth]{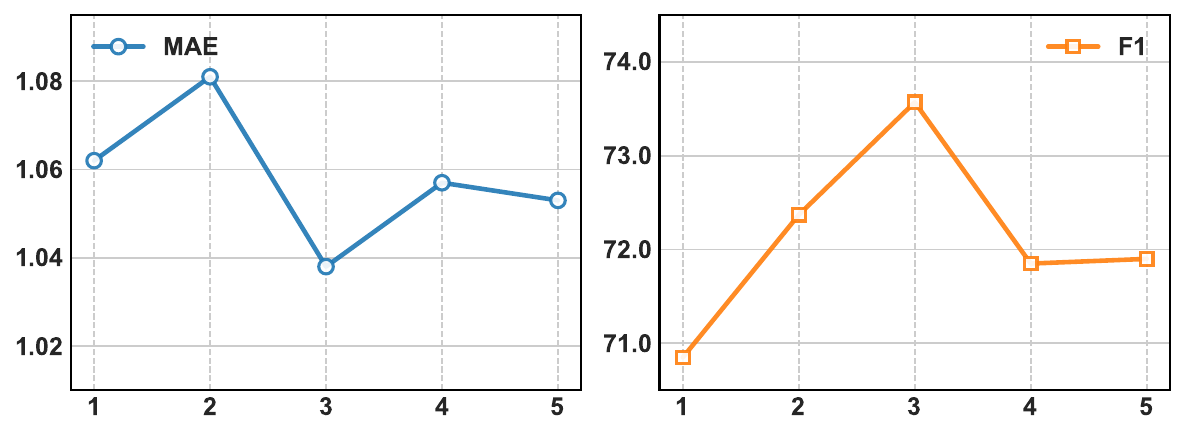}
        \caption{Effect of shared experts}
        \label{fig:sub2}
    \end{subfigure}
\caption{Sensitivity analysis of the number of private and shared experts on model performance on MOSI.}  
\label{fig:sensitivity}
\end{figure}

\begin{figure}[h]
    \centering
    \begin{subfigure}[b]{\textwidth}
        \centering
        \includegraphics[width=\textwidth]{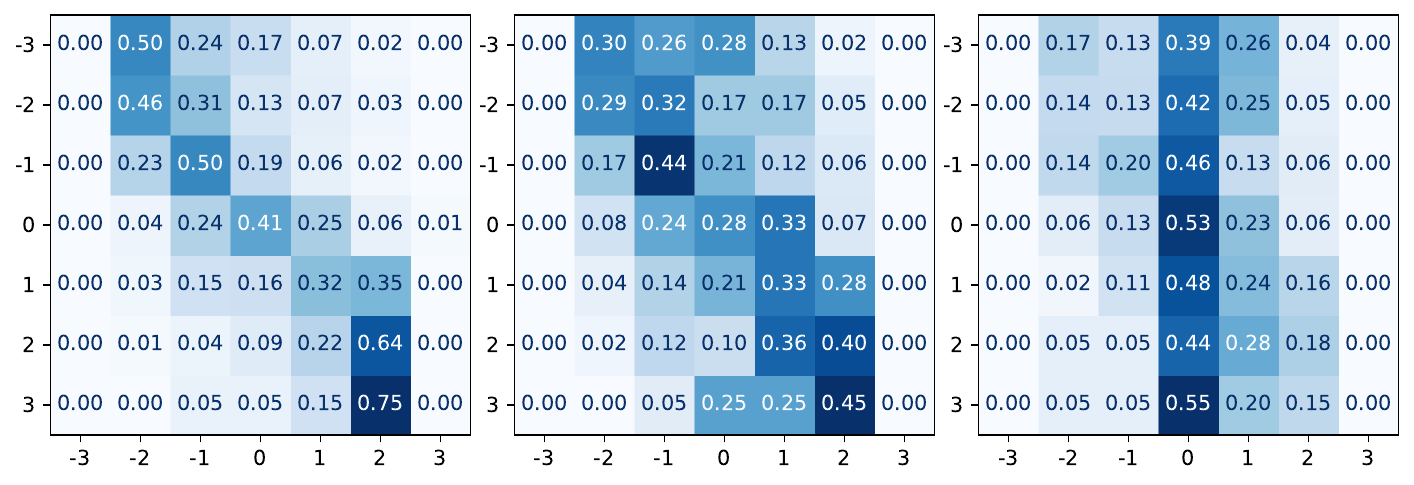}
        \caption{Seven-class confusion matrices of DMD}
        \label{fig:dmd}
    \end{subfigure}
    \hfill
    \begin{subfigure}[b]{\textwidth}
        \centering
        \includegraphics[width=\textwidth]{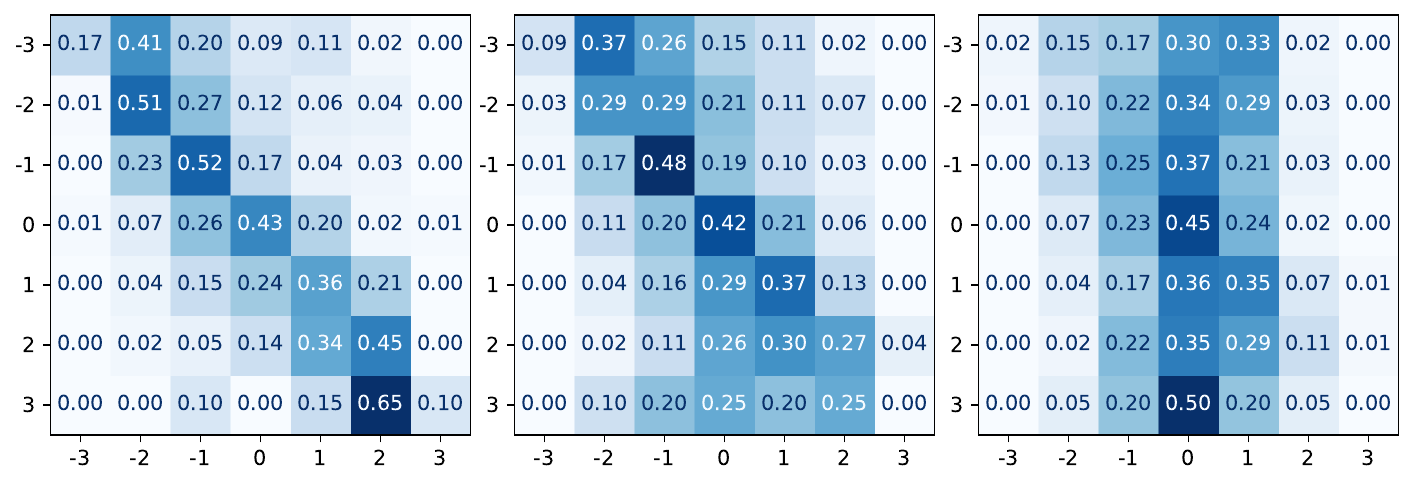}
        \caption{Seven-class confusion matrices of DERL}
        \label{fig:derl}
    \end{subfigure}
    \caption{Seven-class confusion matrices of DERL and DMD under different missing rates on MOSI. The results from left to right correspond to $r = 0.1$, $r = 0.5$, and $r = 0.9$.}
    \label{fig:confusion}
\end{figure}
\subsection{Further Analysis}
\paragraph{Effect of Missing Rates}
To investigate the impact of the predefined missing rate on model performance, we evaluate the model under different missing rates $r$. The results, summarized in Table~\ref{tab:missing_rate_sensitivity}, illustrate the  robustness across varying levels of incompleteness.
As the missing rate increases, the performance first improves and then declines. At low missing levels, the gap between incomplete and complete inputs is minor,  which limits its ability to learn informative representations from missing modalities. Conversely, excessively high missing rates severely degrade input quality, making it difficult for the model to disentangle and reconstruct meaningful sentiment cues.
A moderate missing rate provides the best trade-off, encouraging the model to learn more robust and balanced cross-modal representations while avoiding over-degradation of model performance. 

\paragraph{Sensitivity of Expert Numbers}
We perform a sensitivity analysis to investigate how the number of mixture experts affects model performance. The numbers of private and shared experts are varied from 1 to 5, and the model is evaluated on the MOSI dataset under intra-modality missing conditions.
Figure~\ref{fig:sensitivity} illustrates the influence of varying private and shared expert numbers on key metrics, including MAE and F1. We observe that when the number of private experts exceeds four, performance declines noticeably, likely due to overfitting. Increasing the number of shared experts enhances cross-modal correlation learning and improves robustness under missing-modality scenarios. However, an excessive number of shared experts introduces  information redundancy and reduces robustness.
To sum up, too few experts constrain the model’s ability to disentangle heterogeneous representations, while too many increase the risk of overfitting and computational burden. The best performance on MOSI is achieved when private and shared experts are balanced at moderate levels (e.g., one private and three shared experts), effectively capturing both modality-specific and shared semantic information.

\paragraph{Visualization of Confusion Matrices}
To further assess the effectiveness of the proposed DERL framework, we compare it with the feature-disentangled baseline DMD on the MOSI dataset. Under different predefined missing rates ($r = 0.1, 0.5, 0.9$), we visualize the confusion matrices of both models on the challenging seven-class sentiment classification task. As shown in Figure~\ref{fig:confusion}, the performance of both models declines as the missing rate increases, while their difference becomes more evident.
At a low missing rate ($0.1$), both models perform similarly, with DERL showing a clearer diagonal pattern and stronger class separability. When the missing rate increases to $0.5$, DERL maintains a consistent advantage. Under the extreme setting ($0.9$), the contrast is most pronounced: DMD exhibits a lazy phenomenon where most predictions concentrate around the neutral class. In contrast, DERL, though slightly affected by missing information, maintains a more balanced prediction distribution and preserves sensitivity to extreme sentiment classes ($-3$ and $3$).
Besides greater robustness, DERL achieves competitive efficiency compared to DMD (see Table~\ref{tab:efficiency_mosi}), effectively capturing sentiment-related information under uncertain missing modalities.
\begin{table}[!h]
    \centering
    \caption{Efficiency comparison on  MOSI under intra-modal missing conditions. The values in parentheses denote the total parameters excluding the text BERT encoder.}
    \resizebox{\linewidth}{!}{
    \begin{tabular}{cccccc}
        \toprule   
        {Model} & Params (M) & Time/Epoch (S) & MAE  & F1 & Acc-7  \\ 
        \midrule
        TFR-Net  & 124.33 (14.85) & 11.53 & 1.200  & 60.06 / 61.73 & 29.54 \\
        DMD  & 122.06 (12.58) & 35.97 & 1.097 & 71.15 / 72.03 &  33.25 \\
        LNLN  & 115.97 (6.48) & 11.11 & 1.062 & 70.02 / 71.71 &  32.53 \\
        \midrule
        \textbf{DERL}  & \textbf{112.38} (\textbf{2.89}) & \textbf{7.29} & \textbf{1.038} & \textbf{71.95} / \textbf{73.57} &  \textbf{34.50} \\
        \bottomrule
    \end{tabular}
    }
    \label{tab:efficiency_mosi}
\end{table}
\subsection{Parameter Efficiency Analysis}
Table~\ref{tab:efficiency_mosi} compares DERL with representative robust multimodal sentiment analysis methods on MOSI in terms of model size, training efficiency, and performance. The values in parentheses indicate parameter counts after removing the text BERT encoder, enabling a fair comparison of multimodal modeling complexity.
DERL consistently maintains a smaller parameter footprint under both full and BERT-free settings. Compared to TFR-Net and LNLN, DERL reduces redundant cross-modal interactions through structurally disentangled expert modeling. Relative to DMD, DERL further compresses modality interaction parameters, retaining only 2.89M core parameters without the text encoder.
In terms of computational efficiency, DERL achieves the fastest training speed among all compared methods, with a per-epoch training time of 7.29 seconds, while still delivering the best or competitive performance on key metrics such as Acc-7, F1, and MAE. This demonstrates that DERL effectively balances efficiency and performance with a compact and principled design.

\section{Conclusion}
\label{sec:conclusion}
In this paper, we present a novel Disentangled Expert Representation Learning (DERL) framework for robust MSA, capable of handling heterogeneous information variations under uncertain missingness. 
Specifically, we propose a hybrid expert disentanglement strategy that adaptively decouples private and shared representations from incomplete inputs.
Moreover, a multi-level feature reconstruction mechanism is introduced to recover missing semantic cues through collaborative supervision. 
Finally, a modality routing and fusion scheme emphasizes dominant modalities while leveraging others to enable importance-aware multimodal fusion.
Extensive experiments under various missing conditions demonstrate the effectiveness of the proposed method.
A limitation of DERL is that it focuses on predefined missing cases, and extending it to more complex missing patterns in real-world scenarios remains challenging.

\section*{Acknowledgments}
This work was supported by the National Natural Science Foundation of China (Nos. 62272025 and U22B2021), and by the Fund of the State Key Laboratory of Software Development Environment.

%
\appendix
\section{Feature Extraction}
\label{sec:app1}
Detailed statistics of MOSI and MOSEI are reported in Table~\ref{tab:dataset_stats}. The feature extraction toolkits are detailed below.
\paragraph{Text Modality} For  MOSI and MOSEI, raw utterances are encoded using the \texttt{bert-base-uncased} model~\cite{devlin2019bert}. The resulting text features have 768 dimensions, with input sequence lengths set to 50 for both datasets.

\paragraph{Audio Modality} Low-level acoustic features are extracted via COVAREP~\cite{degottex2014covarep}, including pitch, MFCC, etc. The audio feature dimensions are 5 for MOSI and 74 for MOSEI, with sequence lengths of 375 and 500, respectively.

\paragraph{Visual Modality} Facial features, encompassing facial action units and head pose, are obtained using the Facet system~\cite{imotions2017fea}. Feature dimensions are 20 for MOSI and 35 for MOSEI, with sequence lengths of 500 for both datasets.
\begin{table}[h]
\caption{For each data split, the three figures correspond to the number of negative ($<$0), neutral ($=$0), and positive ($>$0) samples.}
\label{tab:dataset_stats}
\begin{tabular}{lrrrr}
\toprule
Dataset     & \#Train       & \#Valid         & \#Test      & \#Total \\
\midrule
MOSI    & 552/53/679       & 92/13/124        & 379/30/277     & 2199   \\
MOSEI   & 4738/3540/8048   & 506/433/932      & 1350/1025/2284 & 22856  \\
\bottomrule
\end{tabular}
\end{table}
\section{Baseline Models}
\label{sec:app2}
We compare our model with several representative state-of-the-art methods, including both the complete-modality and missing-modality approaches.   

\textbf{MISA}~\cite{hazarika2020misa} decomposes each modality into shared and private latent subspaces, effectively modeling both cross-modal commonalities and modality-specific properties.

\textbf{Self-MM}~\cite{yu2021learning} employs self-supervised objectives for multimodal and unimodal learning, capturing shared inter-modal patterns while preserving modality uniqueness.

\textbf{MMIM}~\cite{han2021improving} maximizes mutual information across modalities to learn correlated and discriminative representations for enhanced fusion.

\textbf{CENet}~\cite{wang2022cross} reinforces textual embeddings by integrating visual and acoustic cues within a language-centered framework.

\textbf{CubeMLP}~\cite{sun2022cubemlp} enables efficient multimodal communication and fusion through MLP-based interaction blocks.

\textbf{ALMT}~\cite{zhang2023learning} guides the fusion process with textual cues, suppressing irrelevant or conflicting signals from visual and acoustic modalities.

\textbf{DMD}~\cite{li2023decoupled} introduces decoupled multimodal distillation to achieve flexible and adaptive cross-modal knowledge transfer while enhancing unimodal features.

\textbf{EMOE}~\cite{fang2025emoe} dynamically adjusts modality contributions via a mixture-of-experts mechanism and maintains unimodal discriminability through knowledge distillation.

\textbf{MCTN}~\cite{pham2019found} learns robust joint embeddings via inter-modal transformations, employing cycle-consistency loss to preserve information across modalities.

\textbf{TransM}~\cite{wang2020transmodality} leverages Transformer-based inter-modal transformations to jointly encode emotional information from multiple modalities.

\textbf{TFR-Net}~\cite{yuan2021transformer} adopts the Transformer-driven fusion and reconstruction paradigm to improve sentiment prediction under incomplete modality settings.

\textbf{SMIL}~\cite{ma2021smil} applies Bayesian meta-learning to perturb latent spaces, enabling unimodal embeddings to approximate the full-modality representations in missing conditions.

\textbf{GCNet}~\cite{lian2023gcnet} captures speaker-level and temporal dependencies in dialogue through dual graph neural modules, jointly optimizing classification and reconstruction.

\textbf{CorrKD}~\cite{li2024correlation} learns reliable multimodal representations based on cross-modal correlation and knowledge distillation, mitigating the impact of uncertain missing modalities.

\textbf{LNLN}~\cite{zhang2024towards} treats language as the dominant modality and applies dominance correction and multimodal alignment to enhance robustness under missing scenarios.

\bibliographystyle{elsarticle-num} 
\bibliography{ref}






\end{document}